\documentstyle[12pt,epsf]{article}								 
\def\CellGroup{\bgroup}
\def\endCellGroup{\egroup}
\begin{document} 
\vspace*{3cm}
\begin{center}
{\large\bf Muon transverse polarization in the $K_{l2\gamma}$ process
within the SM and three Higgs doublets Weinberg model}
\end{center}

\vspace*{1cm}
\begin{center}
Braguta V.V.$^{\dagger}$, Likhoded A.A.$^{\dagger\dagger}$, Chalov
A.E.$^{\dagger}$
\end{center}

\vspace*{1cm}
\begin{center}
$^{\dagger}$ {\it Moscow Institute of Physics and Technology, Moscow, 141700  Russia}
\\
$^{\dagger\dagger}$ {\it Institute for High Energy Physics, Protvino, 142284 Russia}
\end{center}

\vspace*{2cm}
\underline{\bf Abstract}

\vspace*{0.5cm}
\noindent
We consider two possible sources of the $T$-violating muon transverse
polarization
in the process $K^+ \to \mu^+ \nu \gamma$:
electromagnetic final state interaction in the SM and
contribution  due to charged-Higgs exchange diagrams in the 
framework of the Weinberg three doublets model.
It is shown that at the one-loop level of the SM 
the muon transverse polarization,  $P_T^{SM}$, varies 
within  $(0.0 \div 1.1 \cdot 10^{-3})$ in the Dalitz plot region.
Averaged value of the muon  polarization, $\langle P^{SM}_T \rangle$, 
in the kinematic region of $E_\gamma \geq 20$~MeV is equal to $4.76\cdot
10^{-4}$.
In the case of the model with three Higgs doublets the muon 
transverse polarization is calculated as a function
of charged Higgs masses, imaginary part of the Yukawa coupling 
constants product and vacuum expectation values of the Higgs doublets.
The averaged value of the transverse polarization 
in the Weinberg model  is 
$\langle P^{Higgs}_T \rangle = - 6.62 \cdot 10^{-5}$.
Perspectives to probe the effect caused by
the charged Higgs exchange diagrams in planned kaon experiments
are discussed.

\newpage
\section{Introduction}

\noindent
The study of the radiative $K$-meson decays is extremely interesting
from the point of view of searching for the new physics effects beyond the Standard
model (SM). One of the most appealing possibilities is to search for
new interactions, which could lead to $CP$-violation.
Contrary to SM, where the $CP$-violation is caused by the presence of the complex phase 
in the  CKM matrix,  the $CP$-violation in extended models, for instance, in models
with three and more Higgs doublets can naturally arises due to the complex couplings
of new Higgs bosons to fermions [1]. Such effects can be probed
in experimental observables, which are essentially sensitive to $T$-odd contributions. 
These observables, for instance, are $T$-odd correlation ($T=\frac{1}{M_K^3}
\vec p_{\gamma}\cdot [\vec p_{\pi} \times \vec p_l ]$) in 
$K^{\pm}\to\pi^0 \mu^{\pm}\nu\gamma$ decay [2] and muon transverse polarization in 
$K^{\pm}\to \mu^{\pm}\nu\gamma$ one.
The search for new physics effects using  the $T$-odd correlation analysis
in the $K^{\pm}\to\pi^0 \mu^{\pm}\nu\gamma$  decay 
will be done in the proposed OKA experiment [3],
where the statistics of $7.0\cdot 10^5$ events for the 
$K^+\to\pi^0 \mu^+\nu\gamma$ decay is expected.

At the moment the  E246 experiment at KEK [4] performs the $K^{\pm}\to \mu^{\pm}\nu\gamma$ 
data analysis searching for $T$-violating muon transverse polarization.
It should be noted that expected value of new physics contribution to the  $P_T$
can be of the order of 
$\simeq 7.0\cdot 10^{-3}\div 6.0\cdot 10^{-2}$ [5,6], 
depending on the extended model type.
Thus, looking for the new physics effects in the muon transverse polarization
it is extremely important to estimate the contribution from so called 
``fake'' polarization, which is caused by the SM electromagnetic final state
interactions and which is natural background for new interaction contributions. 

The Weinberg model with three Higgs doublets [1,6] is especially interesting
for the search of possible $T$-violation. This model allows one to have 
complex Yukawa couplings that leads to extremely interesting phenomenology. 
It was shown [2] that the study of the $T$-odd correlation
in the $K^+\to\pi^0 \mu^+\nu\gamma$ process allows one either to probe the terms, which 
are linear in $CP$-violating couplings, or put the strict bounds on the Weinberg model
parameters. 
So, it seems important to analyze possible effects, which this model can induce
in the muon transverse polarization in the $K^{\pm}\to \mu^{\pm}\nu\gamma$ decay, as well.

In this paper we investigate two possible sources of the muon transverse polarization
in the  $K^{\pm}\to \mu^{\pm}\nu\gamma$ process: а) the effect, induced by the
electromagnetic final state interaction in the one-loop approximation
of the minimal Quantum Electrodynamics, б) the effect, induced by the charged-Higgs
exchange within the three-doublets Weinberg model.

 In next Section we present the calculations of the muon transverse polarization
 with account for one-loop diagrams with final state interactions within the SM. 
 In Section 3 we calculate the muon transverse polarization caused by the 
diagrams with charged-Higgs exchange, where new charged Higgs bosons have 
complex couplings to fermions. Last Section summarized  the results and 
conclusions.

\section{Muon transverse polarization in the  $K^+ \to \mu^+ \nu \gamma$ process
within SM}

\noindent
The $K^+ \rightarrow \mu^+ \nu \gamma$ decay in the tree level of SM
is described by the diagrams shown in  Fig.~1. The diagrams in Fig. 1b and 1c
correspond to the muon and kaon bremsstrahlung, while the diagram in Fig. 1a 
corresponds to the structure radiation.
This decay amplitude can be written as follows
\begin{equation}
M=ie \frac{G_F}{\sqrt{2}}V^{*}_{us}\varepsilon^{*}_{\mu}\left(f_K m_{\mu}
\overline u(p_{\nu})(1+\gamma_{5}) \biggl( \frac{p_{K}^{\mu}}{(p_K q)}
-\frac{(p_{\mu})^{\mu}}{(p_{\mu} q)}-
\frac{\hat{q} \gamma^{\mu}}{2( p_{\mu} q)} \biggr)v(p_{\mu})-G^{\mu \nu}
l_{\nu}
\right) \,
\end{equation}
where 
\begin{eqnarray}
l_{\mu}=\overline u(p_{\nu}) (1+\gamma_{5}) \gamma_{\mu}
v(p_{\mu}) \hspace{1.47in} \nonumber \\
G^{\mu \nu}= i F_v ~\varepsilon^{ \mu \nu \alpha \beta } q_{\alpha}
(p_K)_{\beta} -
F_a ~ ( g^{\mu \nu} (p_K q)-p_K^{\mu} q^{\nu} )\, 
\end{eqnarray}
$G_F$ is the Fermi constant, $V_{us}$ is the corresponding element of the CKM
matrix; $f_K$ is the $K$-meson leptonic constant;
$p_K$, $p_\mu$, $p_\nu$, $q$ are the kaon, muon, neutrino, and photon four-momenta,
correspondingly; $\varepsilon_{\mu}$ is the photon polarization vector; $F_v$ and $F_a$
are the kaon vector and axial formfactors. In formula (2), we use the following
definition of Levi-Civita tensor: $\epsilon^{0 1 2 3} = +1$.

The part of the amplitude, which corresponds to the structure radiation 
and kaon bremsstrahlung and which will be used further in one-loop calculations,
has the form:
\begin{equation}
M_K=ie \frac{G_F}{\sqrt{2}}V^{*}_{us}\varepsilon^{*}_{\mu}\left(f_K m_{\mu}
\overline u(p_{\nu})(1+\gamma_{5}) \biggl( \frac{p_{K}^{\mu}}{(p_K q)} -
\frac {\gamma^\mu}  m_\mu  
  \biggr)v(p_{\mu})-G^{\mu \nu} l_{\nu}
\right)
\end{equation}

The partial width of the $K^+ \to \mu^+ \nu \gamma$ decay in the  
$K$-meson rest frame can be expressed as 
\begin{equation}
d \Gamma = \frac {\sum |M|^2} { 2 m_K } (2 \pi)^4 \delta ( p_K - p_\mu -q
-p_\nu)
\frac {d^3 q} {(2 \pi)^3 2 E_q } \frac {d^3 p_\mu} {(2 \pi)^3 2 E_\mu }
\frac {d^3 p_\nu} {(2 \pi)^3 2 E_\nu }\;,
\end{equation}
where summation over muon and photon spin states is performed.

Introducing the unit vector along the muon spin direction in muon rest frame,
$\bf \vec s $, where ${\bf \vec e}_i~(i=L,\: N,\: T)$ are unit vectors
along longitudinal, normal and transverse components of muon polarization, 
one can write down the squared matrix element of the transition into the particular
muon polarization state in the following form:
\begin{equation}
|M|^2=\rho_{0}[1+(P_L {\bf \vec e}_L+P_N {\bf \vec e}_N+P_T {\bf \vec
e}_T)\cdot \bf \vec s]\;,
\end{equation}
where $\rho_{0}$ is the Dalitz plot probability density averaged over polarization states.
The ${\bf \vec e}_i$ unit vectors can be expressed in term of three-momenta of final 
particles:
\begin{equation}
{\bf \vec e}_L=\frac{\vec p_{\mu}}{|\vec p_{\mu}|}~~~ 
{\bf \vec e}_N=\frac{\vec p_{\mu}\times(\vec q \times \vec p_{\mu})}
{|\vec p_{\mu}\times(\vec q \times \vec p_{\mu})|}~~~
{\bf \vec e}_T=\frac{\vec q \times \vec p_{\mu}}{|\vec q \times \vec
p_{\mu}|}\: . 
\end{equation}
With such definition of ${\bf \vec e}_i$ vectors, $P_T, P_L$, and $P_N$ denote
transverse, longitudinal, and normal components of the muon polarization correspondingly.
It is convenient to use the following variables
\begin{equation}
x=\frac{2 E_\gamma}{m_K}~~~y=\frac{2 E_\mu}{m_K}
~~~\lambda=\frac{x+y-1-r_\mu}{x}~~~r_{\mu}=\frac {m_{\mu}^2}{m_K^2}\;,
\end{equation}
where $E_\gamma$ and $E_\mu$ are the photon and muon energies in the kaon
rest frame. Then, the Dalitz plot probability density, as a function of the  $x$ and $y$
variables, has the form:
\begin{eqnarray}
\rho_{0}=\frac{1}{2}e^2 G_F^2 |V_{us}|^2 &\cdot & \biggl( 
\frac {4 m_{\mu}^2 |f_K|^2 } {\lambda x^2} (1-\lambda) \Bigl (x^2+2 (1-r_{\mu})
(1-x-\frac {r_{\mu}} {\lambda} )\Bigr)	\nonumber\\
&+& m_K^6 x^2 (|F_a|^2+|F_v|^2) (y-2 \lambda y-\lambda x +2
\lambda^2)\nonumber\\
&+& 4  ~\hbox{Re} (f_K F_v^*) ~m_K^4 r_{\mu} \frac x {\lambda} (\lambda -1)
\hspace{0.3in} \nonumber\\
&+&4 ~\hbox{Re} (f_K F_a^*) ~m_K^4  r_{\mu} (-2 y+x+2 \frac {r_{\mu}} {\lambda}
-
\frac x {\lambda} +2 \lambda )\hspace{0.3in} \nonumber\\
&+& 2 ~\hbox{Re} (F_a F_v^*) ~m_K^6 x^2 (y-2 \lambda +x \lambda) \biggr)
\end{eqnarray}
Calculating the muon transverse polarization $P_T$ we follow the original work [7]
and assume that the amplitude of the decay is $CP$-invariant,
and the $ f_K $, $ F_v$, and $ F_a$ formfactors are real.
In this case the tree level muon polarization $P_T=0$.
When one-loop contributions are incorporated, the nonvanishing muon
transverse polarization can arise due to the interference of tree-level 
diagrams and imaginary parts of one-loop diagrams, induced by the electromagnetic 
final state interaction.

To calculate formfactor imaginary parts one can use the
$S $-matrix unitarity:
\begin{equation}
S^+ S=1
\end{equation}
and, using $ S=1+i T$, one gets:
\begin{equation}
T_{f i}-T_{i f}^*=i \sum_n T^*_{n f} T_{n i}, 
\end{equation}
where $i,\: f, \: n$ indices correspond to the initial, final, and intermediate
states of the particle system.
Further, using the  $T$-invariance of the matrix element one gets:
\begin{eqnarray}
\mbox{Im} T_{f i}=\frac 1 2 \sum_n T^*_{n f} T_{n i}  \\
T_{f i}=(2 \pi)^4 \delta ( P_f-P_i ) M_{f i}
\end{eqnarray}
SM one-loop diagrams, which contribute to the muon transverse polarization in
the $K^+ \to \mu^+ \nu \gamma$ decay, are shown in Fig. 2.
Using Eq. (3) one can write down imaginary parts of these diagrams.
For diagrams in Figs. 2a, 2c one has
\begin{eqnarray}
~\hbox{Im}M_1=\frac{i e \alpha}{2 \pi}\frac{G_F}{\sqrt{2}}V_{us}^*
\overline u(p_{\nu})(1+\gamma_{5}) \int \frac{d^3 k_{\gamma}}{2
\omega_{\gamma}}
\frac{d^3 k_{\mu}}{2 \omega_{\mu}}\delta(k_{\gamma}+k_{\mu}-P)R_{\mu}\times
\nonumber\\
(\hat k_{\mu}-m_{\mu})\gamma^{\mu}\frac{\hat q+ \hat
p_{\mu}-m_{\mu}}{(q+p_{\mu})^2-m_{\mu}^2}
\gamma^{\delta} \varepsilon^*_{\delta} v(p_{\mu}) \label{im1}
\end{eqnarray}
For diagrams in Figs. 2b, 2d one has
\begin{eqnarray}
~\hbox{Im}M_2=\frac{i e \alpha}{2 \pi}\frac{G_F}{\sqrt{2}}V_{us}^*
\overline u(p_{\nu})(1+\gamma_{5}) \int \frac{d^3 k_{\gamma}}{2
\omega_{\gamma}}
\frac{d^3 k_{\mu}}{2 \omega_{\mu}}\delta(k_{\gamma}+k_{\mu}-P)R_{\mu}\times
\nonumber\\
(\hat k_{\mu}-m_{\mu})\gamma^{\delta} \varepsilon^*_{\delta} 
\frac{ \hat k_{\mu}-\hat q-m_{\mu}}{(k_{\mu}-q)^2-m_{\mu}^2}
\gamma^{\mu} v(p_{\mu}), \label{im2}
\end{eqnarray}
where
\begin{eqnarray}
R_{\mu}&=&f_K m_{\mu} \biggl (\frac {(p_K)_{\mu}} {(p_K k_{\gamma})} -
\frac {\gamma_{\mu}}{m_{\mu}} \biggr) -i F_v \varepsilon_{\mu \nu \alpha \beta} 
(k_{\gamma})^{\alpha} (p_K)^{\beta} \gamma^{\nu}\nonumber \\
& + & F_a (\gamma_{\mu} (p_K k_{\gamma})-(p_K)_{\mu} \hat k_{\gamma})\:.
\end{eqnarray}

To write down the contributions from diagrams in Figs. 2e, 2f,
$R_{\mu}$ should be substituted by 
\begin{eqnarray}
R_{\mu}&=&f_K m_{\mu} \biggl (
\frac {\gamma_\mu}{m_\mu} - \frac{(k_\mu )_\mu}{(k_\mu k_\gamma )}
-\frac{\hat k_\gamma \gamma_\mu }{2(k_\mu k_\gamma )}\biggr) 
\end{eqnarray}
in expressions (\ref{im1}), (\ref{im2}).

We do not present the expression for the imaginary part
of the diagram in Fig. 2g and further, in calculations,
we neglect this diagram contribution to muon transverse polarization, because, as it will
be shown later, its contribution is negligibly small in comparison with the contribution
from other diagrams. The contribution from this SM diagram was calculated for the first
time in [18], where the authors clarified as well the maximal value of
$P_T$ in generic SUSY models with $R$-parity conservation.

The details of the integrals calculation entering Eqs. (13), (14), and their dependence
on kinematical parameters are given in Appendix~1.

The expression for the amplitude with account for imaginary one-loop contributions
can be written as:
\begin{eqnarray}
M=ie \frac{G_F}{\sqrt{2}}V^{*}_{us}\varepsilon^{*}_{\mu}\Biggl(\tilde f_K
m_{\mu}
\overline u(p_{\nu})(1+\gamma_{5}) \biggl( \frac{p_{K}^{\mu}}{(p_K q)}
-\frac{(p_{\mu})^{\mu}}{(p_{\mu} q)}
\biggr)v(p_{\mu})+ \nonumber\\
\tilde F_n \overline u(p_{\nu})(1+\gamma_{5}) \hat{q} \gamma^{\mu} v(p_{\mu})
-\tilde G^{\mu \nu} l_{\nu}\Biggr),\hspace{1.in}
\end{eqnarray}
where 
\begin{equation}
\tilde G^{\mu \nu}= i \tilde F_v ~\varepsilon^{ \mu \nu \alpha \beta }
q_{\alpha} (p_K)_{\beta} -
\tilde F_a ~ ( g^{\mu \nu} (p_K q)-p_K^{\mu} q^{\nu} ) \;.
\end{equation}
The $\tilde f_K ,\;  \tilde F_v ,\; \tilde F_a$, and 
$\tilde F_n$ formfactors include one-loop contributions 
from diagrams shown in Figs.~2a-2f.  
The choice of the formfactors is determined by the matrix element expansion
into set of gauge-invariant structures.

As far as we are interested in the contributions of 
imaginary parts of one-loop diagrams only, since namely they lead to
nonvanishing transverse polarization, we neglect the real parts of these 
diagrams and assume that $\mbox{Re} \tilde f_K ,\; \mbox{Re} \tilde F_v ,\; \mbox{Re}
\tilde F_a $ coincide with their tree-level values,
$f_K ,\; F_v ,\;  F_a $, correspondingly, and  
$ \mbox{Re} \tilde F_n= - f_K m_\mu /2(p_\mu q)$.
Explicit expressions for  imaginary parts of the formfactors
are given in Appendix~2.

The muon transverse polarization can be written as
\begin{equation}
P_T=\frac {\rho_T} {\rho_0},
\end{equation}
where
\begin{eqnarray}
\rho_T=2 m_K^3 e^2 G_F^2 |V_{us}|^2 x \sqrt{\lambda y - \lambda^2 - r_{\mu}}
\biggl(
m_{\mu} ~\hbox{Im}(\tilde f_K \tilde F_a^*)(1-\frac{2}{x}+
\frac{y}{\lambda x}) \nonumber\\ 	
+ m_{\mu} ~\hbox{Im}(\tilde f_K \tilde F_v^*)(\frac{y}{\lambda x}-1-2
\frac{r_{\mu}}{\lambda x})+
2 \frac{r_{\mu}}{\lambda x} ~\hbox{Im}(\tilde f_K \tilde F_n^*)(1-\lambda)
\hspace {0.4in}\nonumber\\  
+ m_K^2 x ~\hbox{Im}(\tilde F_n \tilde F_a^*)(\lambda-1)+ m_K^2 x ~\hbox{Im}
(\tilde F_n \tilde F_v^*)(\lambda-1)\biggr)		
\hspace{0.6in}
\end{eqnarray}

It should be noted that expression  (20) 
disagrees with $\rho_T$ in [15]. In particular, 
the terms containing Im$F_n$ are missed in the 
$\rho_T$ expression given in [15].
Moreover, calculating the muon transverse polarization we
took into account the diagrams shown in Fig. 2e and 2f, which have been neglected
in [15], and which give the contribution comparable with
the contribution from other diagrams in Fig. 2.

%%%%%%%%%%%%%%%%%%%%%%%%%%%%

\section{Three Higgs doublets Weinberg model}

\vspace*{0.3cm}
\noindent
As it was shown in original paper by Weinberg [1], 
one of possible sources of spontaneous $CP$-violation due to the charged-Higgs
bosons exchange is the presence of different relative phases of vacuum expectation values 
of Higgs doublets.
However, the Natural Flavor Conservation requires at least three Higgs doublets.
In the framework of this model there are three sources of $CP$-violation:
\begin{description}
\item[(I)] the complex CKM-matrix;
\item[(II)] a phase in the charged-Higgs boson mixing;
\item[(III)] neutral scalar-pseudoscalar mixing.
\end{description}
In the original Weinberg three-Higgs-doublet model,
$CP$ is broken spontaneously, CKM-matrix is real,
and observed $CP$-violation in the neutral kaon sector comes solely from charged-Higgs
boson exchange.
However, as it was shown in [9], 
in the framework of this model the charged and neutral Higgs boson exchange 
can lead to the noticeable effect in the NEDM value, while the transverse muon polarization 
in the $K^+ \to \mu^+ \nu \gamma$ decay is affected by 
the charged-Higgs boson exchange only.
So, to single out the effect  {\bf (II)}, 
we will suppose, following the ideology of [6], 
that $CP$-violation effect due to the neutral-Higgs boson exchange is smaller than that one caused by 
the charged-Higgs boson exchange.

In the framework of the Weinberg model the charged Higgs boson interaction
with quarks and leptons and be represented as:
\begin{displaymath}
{\cal L}_Y=(2\sqrt{2}G_F)^{1/2}\sum^2_{i=1}(\alpha_i\bar U_L K
M_DD_R+\beta_i\bar U_R
M_U KD_L
+\gamma_i\bar N_L M_E E_R)H^+_i +H.C.\;,
\end{displaymath}
where $K$ is the CKM matrix, $M_U,\; M_D$ и $M_E$ are the mass matrices 
for quarks of $d$-
and $u$-type and charged leptons, correspondingly; $\alpha_i, \;
\beta_i,\;\gamma_i$ are the complex couplings, with are interrelated as
follows [10]
\begin{displaymath}
\frac{\mbox{Im}(\alpha_2\beta_2^*)}{\mbox{Im}(\alpha_1\beta_1^*)}=
\frac{\mbox{Im}(\beta_2\gamma_2^*)}{\mbox{Im}(\beta_1\gamma_1^*)}=
\frac{\mbox{Im}(\alpha_2\gamma_2^*)}{\mbox{Im}(\alpha_1\gamma_1^*)}=-1
\end{displaymath}
and 
\begin{displaymath}
\frac{1}{v_2^2}\mbox{Im}(\alpha_1\gamma_1^*)=
-\frac{1}{v_1^2}\mbox{Im}(\beta_1\gamma_1^*)=
-\frac{1}{v_3^2}\mbox{Im}(\alpha_1\beta_1^*)
\end{displaymath}
where $v_i \; (i=1,2,3)$ are the vacuum expectation values of
Higgs doublets, $\phi_i$, and
\begin{displaymath}
v=(v_1^2+v_2^2+v_3^2)^{1/2}=(2\sqrt{2}G_F)^{-1/2}
\end{displaymath}

\noindent
In the model with three Higgs doublets the $K^+ \to \mu^+ \nu 
\gamma$ decay amplitude  can be written
as:
\begin{displaymath}
M=M_{SM}+M_{Higgs}\;,
\end{displaymath}
where $M_{SM}$ is the SM part of the amplitude with $W$-boson exchange,
and $M_{Higgs}$ is the part amplitude due to the charged-Higgs boson exchange.
\noindent
The amplitude due to charged-Higgs boson exchange is
\begin{eqnarray}
M_{Higgs}=
-e\frac {G_F}{\sqrt{2}}V^{*}_{us} [ & & \langle\gamma|\overline {s}
\gamma_5 u
|K^+\rangle ~\overline{\nu}(1+\gamma_5)\mu + \nonumber \\
&+&  \langle 0|\overline {s}\gamma_5 u
|K^+\rangle ~\overline{\nu}(1+\gamma_5)\frac{-\hat{p}_\mu-\hat{q}+m_\mu}{2
(p_\mu q)}
\hat{\varepsilon}^*\mu ] J,  \nonumber
\end{eqnarray}
where 
\begin{displaymath}
J=m_{\mu} \sum_{i=1}^2\frac {m_u \beta^{*}_{i}\gamma^{}_{i}-m_s
\alpha^{*}_{i}\gamma^{}_{i}}
{M^{2}_{H_{i}}}\;.
\end{displaymath}
The amplitude gets the contributions from  the 
$\langle 0|\overline{s} \gamma_{5}u|K^+ \rangle$ and 
$\langle\gamma|\overline s \gamma_{5}u|K^+ \rangle$ matrix elements, which
can be expressed in terms of the  $ f_K $ formfactor, using its definition
and requiring the gauge invariance of the matrix element: 
\begin{eqnarray}
\langle 0|\overline{s} \gamma_{5}u|K^+ \rangle &= &-i\frac{m_K^2 f_K}{m_s+m_u}
\nonumber\\
\langle\gamma|\overline s \gamma_{5}u|K^+ \rangle &=&-i\frac{m_K^2
f_K}{(m_s+m_u)(p_K q)}
(p_K \varepsilon^*)\nonumber
\end{eqnarray}			   
From now on, to simplify the expressions, we introduce
the following constant:  
\begin{displaymath}
\eta=\frac{m_K^2 f_K}{m_s+m_u}
\end{displaymath}
Using this notation one can rewrite the total amplitude as  follows
\begin{eqnarray}
M=ie \frac{G_F}{\sqrt{2}}V^{*}_{us}\varepsilon^{*}_{\mu}\biggl(\left(f_K
m_{\mu}-\eta J \right)  
\overline u(p_{\nu})(1+\gamma_{5}) \cdot \hspace*{1.cm} \nonumber \\
\biggl(\frac{p_{K}^{\mu}}{(p_Kq)}
-\frac{(p_{\mu})^{\mu}}{(p_{\mu} q)}-
\frac{\hat{q} \gamma^{\mu}}{2( p_{\mu} q)} \biggr)v(p_{\mu})-
G^{\mu \nu} l_{\nu}\biggr) \nonumber
\end{eqnarray}
Then $\rho_0$ takes the form:
\begin{eqnarray}
\rho_{0}=\frac{1}{2}e^2 G_F^2 |V_{us}|^2 &\cdot & \biggl( 
\frac {4 ( m_{\mu} f_K-\eta J)^2 } {\lambda x^2} (1-\lambda) \Bigl (x^2+2
(1-r_{\mu})
(1-x-\frac {r_{\mu}} {\lambda} )\Bigr)+	 \nonumber\\
&+& m_K^6 x^2 (F_a^2+F_v^2) (y-2 \lambda y-\lambda x +2 \lambda^2)+\nonumber \\
&+& 4  (f_K F_v-\frac{\eta F_v \hbox{Re}(J)}{m_K}) ~m_K^4 r_{\mu} \frac x
{\lambda}
(\lambda -1)+  \nonumber\\
&+&4  (f_K F_a-\frac{\eta F_a \hbox{Re}(J)}{m_K}) ~m_K^4  r_{\mu} (-2 y+x+2
\frac{r_{\mu}} {\lambda} -
\frac x {\lambda} +2 \lambda )+\nonumber \\
&+&2  (F_a F_v) ~m_K^6 x^2 (y-2 \lambda +x \lambda) \biggr) \nonumber
\end{eqnarray}
The muon transverse polarization is $P_T=\rho_T / \rho_0$, where
\begin{eqnarray}
\rho_T = 2 m_K^3 e^2 \eta G_F^2 |V_{us}|^2  \hbox{Im}(J) \cdot \hspace*{8.cm}
\nonumber \\
\sqrt {\lambda y-\lambda^2-r_\mu}
\biggl(
2 x F_v +2 F_a +\frac 1 {\lambda } (2 r_\mu F_v-
(x+y)(F_a+ F_v) )
\biggr) \nonumber
\end{eqnarray}
and\footnote{Here and further we assume that $M_{H_2}>>M_{H_1}\sim
M_H$ and neglect the terms multiplied by $m_u$.}
\begin{equation}
\hbox{Im}(J)= m_{\mu} m_s \frac {\hbox{Im}(\alpha_1 \beta_1^*) } {M_H^2} \frac
{v_2^2} {v_3^2}\;.  \label{imagej}
\end{equation}
Thus, the value of muon transverse polarization in the framework of this 
model is the function of charged-Higgs boson masses, vacuum expectation values and
imaginary part of Yukawa couplings product.
$P_T$ reaches its maximal value at maximal values of
$\mbox{Im}(\alpha_1 \beta_1^*)$ and $v_2^2/v_3^2$ 
and at minimal mass of charged-Higgs boson.
Therefore, to estimate  the model effect in the muon transverse polarization 
one needs to know  bounds on the model parameters, which follow from experimental data. 

As it was pointed out in [6], the bounds on the parameters of Weinberg model can be 
determined from:
\begin{description}
\item[ I.] LEP II data on direct search of charged-Higgs boson.
The current bound [11] on the charged-Higgs boson mass is
\begin{equation}
M_{H^\pm}\geq 69\mbox{ GeV}\;.  \label{bound1}
\end{equation}
\item[ II.] Model bounds for the analog of the CKM-matrix for the charged-Higgs 
boson mixing, which relates vacuum expectation values of Higgs doublets [6]:
\begin{equation}
\frac { v_3 \sqrt {v_1^2+v_2^2+v_3^2}} {2 v_1 v_2} \geq 9
\;.  \label{bound2}
\end{equation}
\item[ III.]
Bounds on the  $v_2/v_1$ ratio from the $ D^0-{\bar D}^0 $-mixing data. Dominant
contribution to the $D$-meson mass difference, $\delta M_D $,
due to the $ D^0-{\bar D}^0 $-mixing and caused by box-diagrams with charged-Higgs bosons,
can written as
\begin{equation}
\delta M_D= \frac{G_F^2}{24 \pi^2 }\sin^2\theta_C m_D f_D^2 
B_D \frac{m_s^4}{M_H^2}(\frac{v_2}{v_1})^4\;. \label{bound3}
\end{equation}
\item[ IV.]
Bounds coming from the experimental data on neutron electric dipole moment
(NEDM). At one-loop level, taking into account 
diagrams with charged-Higgs boson exchange, one can write down the expression
for NEDM as follows [12]:
\begin{eqnarray}
d_{n}=\frac{4}{3}d_{d}&=&-\frac{\sqrt{2}G_{F}}{9
\pi^{2}}m_{d}\mbox{Im}(\alpha_{1}\beta_{1}^{*})
\cdot \nonumber \\
&\cdot & 
\sum_{i=c,t}\frac{x_{i}}{(1-x_{i})^2}\cdot \biggl( \frac{5}{4}x_{i}-
\frac{3}{4}-\frac{1-\frac{3}{2}x_{i}}{1-x_{i}}\ln ~x_{i}\biggr)K_{id}^{2}\;,
\label{bound4}
\end{eqnarray}
где $x_{i}=m_{i}^2/M_{H}^{2}$.
\end{description}

\noindent
In Figs. 3a-3c we present the allowed parameter regions, which follow 
from (\ref{bound1})-(\ref{bound4}). Further, calculating the muon transverse polarization
in the framework of the Weinberg model we will adopt the model parameters taking
into account the bounds above.

\section{Results and discussion}

\noindent
For the numerical calculations we use the following formfactor values
$$ 
f_K=0.16 \mbox{ GeV},\;\; F_v=\frac{0.095}{m_K},\;\; F_a=-\frac{0.043}{m_K}\;.
$$
The $ f_K $ formfactor is determined from experimental data on kaon decays [11],
and $F_v, F_a $ ones are calculated at the one loop-level in the Chiral
Perturbation Theory [13]. It should be noted that our definition for $F_v$
differs by sign from that in [13]. With these choice of formfactor values the 
decay branching with the cut on 
photon energy $E_\gamma \geq 20$~MeV is equal to 
Br($K^{\pm}\to\mu^{\pm}\nu\gamma$) $=3.3\cdot10^{-3}$, that is in good agreement with 
PDG data.

\vspace*{0.5cm}
\underline{\bf The Standard Model case}

\vspace*{0.3cm}
\noindent
The three-dimensional distribution of muon transverse polarization, 
calculated at the SM one-loop approximation is shown in Fig. 4.
It should be noted that $P_T$ as the function of the $x$ and $y$ parameters
is characterized  by the sum of individual contributions of diagrams in Figs. 2a-f, 
while the contributions from diagrams 2a-d [8] and  2e-f are comparable in absolute 
value, but opposite in sign, so the total $P_T (x,y)$ distribution is
the difference of these group contributions and in absolute value
is about one order of magnitude less than each of them.
Our estimates show that in the Dalitz plot region 
the contribution from diagram 2g is an order of magnitude less
than the total contribution from other diagrams.

As one can see in Fig. 4, the maximal absolute value of the 
muon transverse polarization is achieved in two domains of the Dalitz
plot region:

a) at $0.3 \leq x=2E_\gamma /m_K \leq 0.6$ and  $y=2E_\mu /m_K \to 1$; 

б) at $0.5 \leq x=2E_\gamma /m_K \leq 0.7$ and  $0.5 \leq y=2E_\mu /m_K \leq
0.7$.

Indeed, analysing the level lines for $P_T$ , shown in Fig. 5,
it is easy to notice that the maximal values of $P_T$ are located 
near to the $x, y$ values (0.3;1.0) and (0.6;0.6).
It should be noted that  the value of muon transverse polarization
is positive in the whole Dalitz plot region and does not take negative values.
Averaged value of transverse polarization can be obtained by integrating the function
$\rho_T/\Gamma(K^+\to\mu^+\nu\gamma)$
over the physical region, and  with the cut on photon energy $E_\gamma > 20 $~MeV
it is equal to
\begin{equation}
\langle P^{SM}_T \rangle = 4.76 \cdot 10^{-4}\;. \label{resul1}
\end{equation}
%%%%%%%%%%%%%%%%%%%%%%%%%%%%%%%%%%%%

Let us note that obtained numerical value of the averaged  transverse polarization
and $P_T(x,y)$ kinematical dependence in Dalitz plot differ from those given
in [15-17]. As it was calculated in  [15], the $P_T$ value varies in the region
of $(-0.1\div 4.0)\cdot 10^{-3}$ for cuts on the muon and photon energies,
$200<E_\mu<254.5$ MeV, $20<E_\gamma<200$ MeV. We have already mentioned above that

1) The authors of [15]  did not take into account  
terms containing the imaginary part of  the $F_n$ formfactor
(contributing to $\rho_T$),
which are, in general, not small being compared with others.

2) The authors of [15] omitted the diagrams, shown in Fig. 2e, 2f, though, 
as it was mentioned above, their contribution to $P_T$ is comparable with that one
of diagrams in Fig. 2a-2d.

All these points lead to serious disagreement between our results and results
obtained in [15]. In particular, our calculations show that the value of
the muon transverse polarization has negative sign in all Dalitz plot region
and its absolute value varies in the region of $(0.0\div 1.1)\cdot 10^{-3}$, and  
the $P_T$ dependence on the $x, y$ parameters is different than in [15].

We would like to remark that the muon transverse polarization for the same process 
was also calculated in [17], where the contributions from diagrams 2e and 2f
were taken into account. 
However, the calculation method used in [17] does not allow
to compare the analytical results, but as for the numerical 
ones, they differ from our results
and results of [15] as well: though the shape of the destributions 
is similar, $P_T$ value has opposite sign in comparison to ours. 
The absence of the explicit expressions for  $\rho_0$ and $\rho_T$ functions and
imaginary parts of formfactors excludes the possibility to compare results [17] 
with the results by other authors.

\vspace*{0.5cm}
\underline{\bf Weinberg model case}

\vspace*{0.3cm}
\noindent
Calculating the muon transverse polarization within the three doublet model
we choose the model parameters in a way, first, to maximize the
polarization value and, second, to satisfy the bounds of
(\ref{bound1})-(\ref{bound4}). In Fig. 6 we present 
the three-dimensional $P_T$ distribution for the Weinberg model case
with $M_{H^\pm }=70$~GeV, $\mbox{Im} (J) =7 \cdot 10^{-5}$, and kinematical
cut $E_\gamma > 20$~MeV.

The behaviour of the transverse polarization as the function $P^{Higgs}_T =f(x,y)$ in the case 
of the three doublet model is significantly differs from that one in SM. First,
in this model the sign of the transverse polarization
depends on the sign of $\mbox{Im}(J)$, see (\ref{imagej}). Second, the $P^{Higgs}_T =f(x,y)$
function has different behaviour and the region of maximal absolute values
is located in different $(x,y)$ region:
$0.6\leq x\leq 0.8$ and $0.9 \leq y \leq 1.$. 
Corresponding level lines $P^{Higgs}_T =f(x,y)$ are shown in Fig.~7.

Comparing the $P_T$ distributions in the case of SM and Weinberg model, 
Figs. 4 and 6, one can see that the maximal value  of the transverse polarization 
in the case of the Weinberg model is a few times less
than that one in SM. The averaged $P_T$ value, obtained by integrating
over the Dalitz plot region with the cut  $E_\gamma > 20 $~MeV, is
\begin{equation}
\langle P^{Higgs}_T \rangle =- 6.62 \cdot 10^{-5}\;, \label{resul2} 
\end{equation}
that is again an order of magnitude less than (\ref{resul1}). So, for reliable
detection of the charged-Higgs effect one needs the experimental sensitivity
to probe the transverse polarization in the $K_{\mu 2\gamma}$ process 
at the level of $10^{-4}$.
Experiments conducted thus far are sensitive to $P_T$ at the level of 
$1.5\cdot 10^{-2}$ [4], that is evidently insufficient to discover the effect. 

Nevertheless, there are possibilities, connected with the upgrade of the
E246 experiment (expected sensitivity is about $2\cdot 10^{-3}$),
and launch of new experiment E923 [14],
where the usage of a new method of $T$-odd polarization measurement 
allows to achieve the level of $10^{-4}$, 
that seems more optimistic. 
Moreover, comparing the $P_T =f(x,y)$ 
in Figs. 4 и 6, one can notice that the relative $P_T^{Higgs}/P_T^{SM}$
contribution can be significantly enhanced by introducing  cuts in $(x,y)$ region.
The statistics increase will allow one to analyse the distribution for $P_T =f(x,y)$
rather than its average value only.

\section*{Acknowledgements}

\noindent
The authors thank Drs. Kiselev V.V. and Likhoded A.K. for fruitful discussion and
valuable remarks. The authors are also grateful to Bezrukov F., Gorbunov D.
for their remark on icorrect sign of formfactor $F_v$ in our previous results. 
This work is in part supported by the Russian Foundation for Basic Research,
grants 99-02-16558 and 00-15-96645, Russian Education Ministry, grant 
RF~E00-33-062 and CRDF MO-011-0.

\newpage

\normalsize
\vspace*{2cm}
\section*{References}

\vspace*{0.5cm}
\noindent
\begin{description}
\item[1.] S. Weinberg, {\em Phys. Rev. Lett.} {\bf 37} (1976), 651.
\item[2.] A. Likhoded, V. Braguta, A. Chalov, {\em в in press} (see. also
{\bf hep-ex}/0011033).
\item[3.] V.F. Obraztsov and L.G. Landsberg, {\bf hep-ex}/0011033). 
\item[4.] See, for example, {\em Phys. Rev. Lett.} {\bf 83} (1999), 4253;
Yu.G. Kudenko, {\bf hep-ex}/00103007).
\item[5.] J.F. Donoghue and B. Holstein, {\em Phys. Lett.} {\bf B113}(1982),
382;
L. Wolfenstein, {\em Phys. Rev.} {\bf 29} (1984), 2130;
G. Barenboim et al., {\em Phys. Rev.} {\bf 55} (1997), 24213;
M. Kobayashi, T.-T. Lin, Y. Okada, {\em Prog. Theor. Phys.} {\bf 95} (1996),
361; S.S. Gershtein et al., {\em Z. Phys.} {\bf C24} (1984), 305;
R. Garisto, G. Kane, {\em Phys. Rev.} {\bf D44} (1991), 2038.
\item[6.]
G. Belanger, C.Q. Geng, {\em Phys. Rev.} {\bf D44} (1991), 2789.
\item[7.] L.B. Okun and I.B. Khriplovich, {\em Sov. Journ. Nucl. Phys.} {\bf v.6 } (1967), 821.
\item[8.] A. Likhoded, V. Braguta, A. Chalov,  Preprint IHEP 2000-57, 2000;
to appear in {\em Phys. Atom. Nucl.}.
\item[9.] A.R. Zhitnitskii, Sov. J. Nucl. Phys., {\bf 31} (1980), 529;
см. также, C.Q. Geng and J.N. Ng, {\em Phys. Rev.} {\bf  D 42} (1990), 1509.
\item[10.] H.Y. Cheng, {\em Phys. Rev.} {\bf  D 26} (1982), 143.
\item[11.] Review of Particle Physics, {\em Euro. Phys. Journ.} {\bf C15}
(2000). 
\item[12.] G. Beal and N.G. Deshpande, {\em Phys. Lett.} {\bf  B 132} (1983),
427.
\item[13.] J. Bijnens, G. Ecker, J. Gasser, {\em Nucl. Phys.} {\bf B396}
(1993), 81.
\item[14.] M.V. Diwan et al.,  AGS Experiment Proposal 923, 1996.
\item[15.] V.P. Efrosinin, Yu.G. Kudenko, {\em Phys. Atom. Nucl.} {\bf v.62}
(1999), 987.
\item[16.] C.H. Chen, C.Q. Geng, C.C Lih {\bf hep-ph}/9709447
\item[17.] R.N. Rogalyov, {\em Phys.Lett.} {\bf B521} (2001), 243
\item[18.] G. Hiller and G. Isidori, {\em Phys. Lett.}  {B459} (1999), 295.
\end{description}

\newpage
\section*{Appendix 1}			  

\noindent
Calculating the integrals, contributing to (14) and (15),  we use the following
notations:
$$
P=p_\mu+q
$$
$$
d \rho =\frac {d^3 k_{\gamma}} {2 \omega_{\gamma}}
\frac {d^3 k_{\mu}} {2 \omega_{\mu}}\delta(k_\gamma+k_{\mu}-P) 
$$
We present below either the explicit expressions for integrals,
or the set of equations, which being solved, give the parameters,
entering the integrals.
\begin{eqnarray}
J_{11}&=&\int d \rho =\frac  {\pi} 2 \frac {P^2-m_{\mu}^2} {P^2}\;,\nonumber \\
J_{12}&=&\int d \rho \frac 1 {(p_K k_{\gamma})}=
\frac {\pi}	{2 I} \ln \biggl( \frac {(P p_K)+I} {(P p_K)-I} \biggr)\;,\nonumber 
\end{eqnarray}
where
$$
I^2=(P p_K)^2-m_K^2 P^2\;.
$$
$$
\int d \rho \frac {k^{\alpha}_{\gamma}} {(p_K k_\gamma)}=a_{11} p^\alpha_K
+b_{11} P^\alpha \;.
$$
The $a_{11}$ and $b_{11}$ parameters are determined by the following
equation:
\begin{eqnarray}
a_{11}&=&-\frac 1 {(P p_K)^2-m_K^2 P^2}
\biggl( P^2 J_{11}- \frac {J_{12}} 2  (P p_K) (P^2-m_{\mu}^2) \biggr)\;,
\nonumber \\
b_{11}&=&\frac 1 {(P p_K)^2-m_K^2 P^2}
\biggl( (P p_K) J_{11}- \frac {J_{12}} 2  m_K^2 (P^2-m_{\mu}^2)\;.
\biggr)\nonumber
\end{eqnarray}
\begin{eqnarray}
\int d \rho k_{\gamma}^{\alpha} &=& a_{12} P^{\alpha}\;, \nonumber \\
\int d \rho k_{\gamma}^{\alpha} k_{\gamma}^{\beta}&=&a_{13} g^{\alpha \beta}
+b_{13} P^{\alpha} P^{\beta}\;, \nonumber 
\end{eqnarray}
where
\begin{eqnarray}
a_{12}&=&\frac {(P^2-m_{\mu}^2)} {2 P^2} J_{11}\;, \nonumber \\
a_{13}&=&-\frac 1 {12} \frac {(P^2-m_{\mu}^2)^2} {P^2} J_{11}\;, \nonumber \\
b_{13}&=&\frac 1 3 \biggl(\frac {P^2-m_{\mu}^2} {P^2} \biggr)^2 J_{11}\;.
\nonumber 
\end{eqnarray}
\begin{eqnarray}
J_1&=&\int d \rho \frac 1 {(p_K k_\gamma)((p_\mu-k_\gamma)^2-m_\mu^2)}=
-\frac \pi {2 I_1 (P^2-m_\mu^2) } \ln 
\biggl( \frac {(p_K p_\mu)+I_1} {(p_K p_\mu)-I_1} \biggr)\;, \nonumber \\
J_2&=&\int d \rho \frac 1 {(p_\mu-k_\gamma)^2-m_\mu^2}=
-\frac \pi {4 I_2} \ln \biggl( \frac {(P p_\mu)+I_2} {(P p_\mu)-I_2} \biggr)
\;, \nonumber 
\end{eqnarray}
where
\begin{eqnarray}
I_1^2&=&(p_K p_\mu)^2 -m_\mu^2 m_K^2\;, \nonumber\\
I_2^2&=&(P p_\mu)^2-m_\mu^2 P^2\;. \nonumber
\end{eqnarray}
\begin{eqnarray}
\int d \rho \frac {k_\gamma^\alpha} {(p_\mu-k_\gamma)^2-m_\mu^2}&=&
a_1 P^\alpha + b_1 p_\mu^\alpha\;, \nonumber \\
a_1&=&-\frac {m_\mu^2 (P^2-m_\mu^2) J_2+(P p_\mu)J_{11}} {2 ((P
p_\mu)^2-m_\mu^2 P^2)} 
\;, \nonumber\\
b_1&=&\frac {(P p_\mu)(P^2-m_\mu^2) J_2+P^2 J_{11}} {2 ((P p_\mu)^2-m_\mu^2
P^2)}
\;, \nonumber
\end{eqnarray}
The integrals below are determined by the parameters, which can be obtained by solving
the sets of equations.
$$
\int d \rho \frac {k_\gamma^\alpha} {(p_K
k_\gamma)((p_\mu-k_\gamma)^2-m_\mu^2)}=
a_2 P^\alpha + b_2 p_K^\alpha +c_2 p_\mu^\alpha \;,
$$
$$
\left\{
\begin{array}{ccc}
a_2 (P p_K)+ b_2 m_K^2+c_2 (p_K p_\mu)=J_2 \hfill \\
a_2 (P p_\mu)+b_2 (p_K p_\mu)+c_2 m_\mu^2=-\frac 1 2 J_{12} \hfill \\
a_2 P^2+b_2 (P p_K)+c_2 (P p_\mu)=(p_\mu q) J_1 \hfill 
\end{array}
\right.
$$
\begin{eqnarray}
\int d \rho \frac {k_\gamma^\alpha k_\gamma^\beta} 
{(p_K k_\gamma)((p_\mu-k_\gamma)^2-m_\mu^2)}&=&
a_3 g^{\alpha \beta}+b_3 (P^\alpha p_K^\beta+P^\beta p_K^\alpha)+
c_3 (P^\alpha p_\mu^\beta+P^\beta p_\mu^\alpha) \nonumber \\
&+& d_3 (p_K^\alpha p_\mu^\beta+p_K^\beta p_\mu^\alpha)+
e_3 p_\mu^\alpha p_\mu^\beta \nonumber \\
&+&f_3 P^\alpha P^\beta +g_3 p_K^\alpha p_K^\beta \;,\nonumber
\end{eqnarray}
\small
$$
\left\{
\begin{array}{cccccccc}
4 a_3+2 b_3 (P p_K)+2 c_3 (P p_\mu)+2 d_3 (p_K p_\mu)+g_3 m_K^2 + e_3
m_\mu^2+f_3 P^2=0 \hfill \\
c_3 (p_K p_\mu) + b_3 m_K^2 + f_3 (P p_K)-a_1=0 \hfill \\
c_3 (P p_K)+d_3 m_K^2+e_3 (p_K p_\mu)-b_1=0 \hfill \\
a_3 + b_3 (P p_K)+d_3 (p_K p_\mu)+g_3 m_K^2=0 \hfill \\
b_3 (p_K p_\mu)+c_3 m_\mu^2+f_3 (P p_\mu)=- \frac 1 2 b_{11} \hfill \\
b_3 (P p_\mu)+d_3 m_\mu^2+g_3 (p_K p_\mu)=-\frac 1 2 a_{11}	 \hfill \\
a_3 P^2+2 b_3 P^2 (P p_K)+2 c_3 P^2 (P p_\mu)+2 d_3 (P p_\mu) (P p_K)+ 
e_3 (P p_\mu)^2+f_3 (P^2)^2+g_3 (P p_K)^2=(p_\mu q)^2J_1 \hfill \\
\end{array}
\right.
$$
\normalsize
$$
\int d \rho \frac {k_\gamma^\alpha k_\gamma^\beta}
{(p_\mu-k_\gamma)^2-m_\mu^2}=
a_4 g_{\alpha \beta}+b_4 (P^\alpha p_\mu^\beta+P^\beta p_\mu^\alpha)+
c_4 P^\alpha P^\beta +d_4 p_\mu^\alpha p_\mu^\beta \;,
$$
$$
\left\{
\begin{array}{cccc}
a_4+d_4 m_\mu^2+b_4 (P p_\mu)=0 \hfill \\
b_4 m_\mu^2+c_4 (P p_\mu)=-\frac 1 2 a_{12} \hfill \\
4 a_4+2 b_4 (P p_\mu)+c_4 P^2+d_4 m_\mu^2=0 \hfill \\
a_4 P^2+2 b_4 P^2 (P p_\mu)+c_4 (P^2)^2+d_4 (P p_\mu)^2==\frac
{(P^2-m_\mu^2)^2} 4 J_2 
\end{array}
\right.
$$

\bigskip

\begin{eqnarray}
\int d \rho \frac {k_\gamma^\alpha k_\gamma^\beta k_\gamma^\delta}
{(p_\mu-k_\gamma)^2-m_\mu^2}&=&a_5 (g^{\alpha \beta} p_\mu^\delta+
g^{ \delta \alpha} p_\mu^\beta+g^{\beta \delta} p_\mu^\alpha)+
b_5 (g^{\alpha \beta} P^\delta +
g^{ \delta \alpha} P^\beta+g^{\beta \delta} P^\alpha) \nonumber \\
&+& c_5 p_\mu^\alpha  p_\mu^\beta  p_\mu^\delta  
+ d_5 P^\alpha P^\beta P^\delta+
e_5 (P^\alpha p_\mu^\beta p_\mu^\delta+P^\delta p_\mu^\alpha p_\mu^\beta+
P^\beta p_\mu^\delta p_\mu^\alpha)\nonumber \\
&+& f_5 (P^\alpha P^\beta p_\mu^\delta+P^\delta P^\alpha p_\mu^\beta+
P^\beta P^\delta p_\mu^\alpha) \;,\nonumber
\end{eqnarray}
\small
$$
\left\{
\begin{array}{cccccc} 
2 a_5+c_5 m_\mu^2+e_5 (P p_\mu)=0 \hfill \\
a_5 m_\mu^2+b_5 (P p_\mu)=-\frac {1}{2}a_{13} \hfill \\
b_5 + e_5 m_\mu^2+f_5 (P p_\mu)=0 \hfill \\
d_5 (P p_\mu)+f_5 m_\mu^2=-\frac 1 2 b_{13} \hfill \\
6 a_5+c_5 m_\mu^2+2 e_5 (P p_\mu)+f_5 P^2=0 \hfill \\
3 a_5   P^2 (P p_\mu)+3 b_5 ( P^2)^2 +c_5 (P p_\mu)^3+
d_5 (P^2)^3+3 e_5 P^2  (P p_\mu)^2+
3 f_5 (P^2)^2 (P p_\mu)=\frac {(P^2-m_\mu^2)^3} 8 J_2
\end{array}
\right.
$$
\normalsize
\newpage
\section*{Appendix 2}

\noindent
Here we present the expressions for imaginary parts of form-factors
as the  functions of parameters, calculated in Appendix 1.

\begin{eqnarray}
{\hbox{Im}\tilde{f}_K}&=&\frac{\alpha}{2\pi} {f_K}\
\big(-4\ {a_3}\
{(p_Kq)}+4\ {a_2}\ {{{m_{\mu}}}^2}\
{(p_Kq)}-2\ {b_3}\ {{{m_{\mu}}}^2}\
{(p_Kq)}+4\ {c_2}\ {{{m_{\mu}}}^2}\
{(p_Kq)}-  \nonumber \\
&&  4\ {c_3}\ {{{m_{\mu}}}^2}\
{(p_Kq)}-2\ {d_3}\ {{{m_{\mu}}}^2}\
{(p_Kq)}-2\ {e_3}\ {{{m_{\mu}}}^2}\
{(p_Kq)}-2\ {f_3}\ {{{m_{\mu}}}^2}\
{(p_Kq)}+   \nonumber \\
 && 4\ {a_2}\ {(p_Kq)}\ {(p_{\mu}q)}-4\
{b_3}\ {(p_Kq)}\ {(p_{\mu}q)}-4\ {c_3}\
{(p_Kq)}\ {(p_{\mu}q)}-4\ {f_3}\
{(p_Kq)}\ {(p_{\mu}q)}\big)+   \nonumber \\
&&\frac{\alpha}{2\pi} {F_a}\ \big(8\ {a_4}\
{(p_Kq)}-8\
{a_5}\ {(p_Kq)}-8\ {b_5}\ {(p_Kq)}+8\
{b_4}\ {{{m_{\mu}}}^2}\ {(p_Kq)}+   \nonumber \\
&& 4\ {c_4}\ {{{m_{\mu}}}^2}\
{(p_Kq)}-2\ {c_5}\ {{{m_{\mu}}}^2}\
{(p_Kq)}+4\ {d_4}\ {{{m_{\mu}}}^2}\
{(p_Kq)}-2\ {d_5}\ {{{m_{\mu}}}^2}\
{(p_Kq)}-   \nonumber \\
&& 6\ {e_5}\ {{{m_{\mu}}}^2}\
{(p_Kq)}-6\ {f_5}\ {{{m_{\mu}}}^2}\
{(p_Kq)}+12\ {b_4}\ {(p_Kq)}\
{(p_{\mu}q)}+8\ {c_4}\ {(p_Kq)}\
{(p_{\mu}q)}+   \nonumber \\
&& 4\ {d_4}\ {(p_Kq)}\ {(p_{\mu}q)}-4\
{d_5}\ {(p_Kq)}\ {(p_{\mu}q)}-4\ {e_5}\
{(p_Kq)}\ {(p_{\mu}q)}-8\ {f_5}\
{(p_Kq)}\ {(p_{\mu}q)}\big)+   \nonumber \\
&& \frac{\alpha}{2\pi} {F_v}\ \big(8\ {a_4}\
{(p_Kq)}-8\
{a_5}\ {(p_Kq)}-8\ {b_5}\ {(p_Kq)}+8\
{b_4}\ {{{m_{\mu}}}^2}\ {(p_Kq)}+   \nonumber \\
&& 4\ {c_4}\ {{{m_{\mu}}}^2}\
{(p_Kq)}-2\ {c_5}\ {{{m_{\mu}}}^2}\
{(p_Kq)}+4\ {d_4}\ {{{m_{\mu}}}^2}\
{(p_Kq)}-2\ {d_5}\ {{{m_{\mu}}}^2}\
{(p_Kq)}-   \nonumber \\
&& 6\ {e_5}\ {{{m_{\mu}}}^2}\
{(p_Kq)}-6\ {f_5}\ {{{m_{\mu}}}^2}\
{(p_Kq)}+12\ {b_4}\ {(p_Kq)}\
{(p_{\mu}q)}+8\ {c_4}\ {(p_Kq)}\
{(p_{\mu}q)}+   \nonumber \\
&& 4\ {d_4}\ {(p_Kq)}\ {(p_{\mu}q)}-4\
{d_5}\ {(p_Kq)}\ {(p_{\mu}q)}-4\ {e_5}\
{(p_Kq)}\ {(p_{\mu}q)}-8\ {f_5}\
{(p_Kq)}\ {(p_{\mu}q)}\big) \nonumber 
\end{eqnarray}

\begin{eqnarray}
{\hbox{Im}\tilde{F_a}}&=&\frac{\alpha}{2\pi}{f_K}\
\Big({a_2}\
{{{m_{\mu}}}^2}+2\ {c_2}\
{{{m_{\mu}}}^2}-{c_3}\ {{{m_{\mu}}}^2}-2\
{d_3}\ {{{m_{\mu}}}^2}-{e_3}\
{{{m_{\mu}}}^2}- \nonumber \\
&& \frac{{a_1}\
{{{m_{\mu}}}^2}}{{(p_{\mu}q)}}-\frac{{b_1}\
{{{m_{\mu}}}^2}}{{(p_{\mu}q)}}+\frac{2\ {b_4}\
{{{m_{\mu}}}^2}}{{(p_{\mu}q)}}+\frac{{c_4}\
{{{m_{\mu}}}^2}}{{(p_{\mu}q)}}+\frac{{d_4}\
{{{m_{\mu}}}^2}}{{(p_{\mu}q)}}\Big)+ \nonumber  \\
&& \frac{\alpha}{2\pi}{F_v}\ \big(8\ {a_4}-4\
{a_5}-12\
{b_5}-2\ {a_1}\ {{{m_{\mu}}}^2}+4\
{b_4}\ {{{m_{\mu}}}^2}+5\ {c_4}\
{{{m_{\mu}}}^2}-{c_5}\ {{{m_{\mu}}}^2}-  \nonumber \\
&& {d_4}\ {{{m_{\mu}}}^2}-3\ {d_5}\
{{{m_{\mu}}}^2}-5\ {e_5}\ {{{m_{\mu}}}^2}-7\
{f_5}\ {{{m_{\mu}}}^2}+2\ {a_1}\
{(p_Kp_{\mu})}-4\ {b_4}\ {(p_Kp_{\mu})}-  \nonumber \\
&& 4\ {c_4}\ {(p_Kp_{\mu})}+2\ {d_5}\
{(p_Kp_{\mu})}+2\ {e_5}\ {(p_Kp_{\mu})}+4\
{f_5}\ {(p_Kp_{\mu})}+2\ {a_1}\
{(p_Kq)}-  \nonumber \\
&& 2\ {b_4}\ {(p_Kq)}-4\ {c_4}\
{(p_Kq)}+2\ {d_5}\ {(p_Kq)}+2\ {f_5}\
{(p_Kq)}-4\ {a_1}\ {(p_{\mu}q)}+  \nonumber \\
&& 6\ {b_4}\ {(p_{\mu}q)}+10\ {c_4}\
{(p_{\mu}q)}-6\ {d_5}\ {(p_{\mu}q)}-2\
{e_5}\ {(p_{\mu}q)}-8\ {f_5}\
{(p_{\mu}q)}\big)+  \nonumber \\
&& \frac{\alpha}{2\pi}{F_a}\ \big(-6\ {a_4}+2\
{a_5}+{c_4}\ {{{m_{\mu}}}^2}-{d_4}\
{{{m_{\mu}}}^2}-{d_5}\ {{{m_{\mu}}}^2}-  \nonumber \\
&& {e_5}\ {{{m_{\mu}}}^2}-2\ {f_5}\
{{{m_{\mu}}}^2}+2\ {a_1}\ {(p_Kp_{\mu})}-4\
{b_4}\ {(p_Kp_{\mu})}-4\ {c_4}\
{(p_Kp_{\mu})}+  \nonumber \\
&& 2\ {d_5}\ {(p_Kp_{\mu})}+2\ {e_5}\
{(p_Kp_{\mu})}+4\ {f_5}\ {(p_Kp_{\mu})}+2\
{a_1}\ {(p_Kq)}-2\ {b_4}\ {(p_Kq)}- \nonumber  \\
&& 4\ {c_4}\ {(p_Kq)}+2\ {d_5}\
{(p_Kq)}+2\ {f_5}\ {(p_Kq)}+2\ {c_4}\
{(p_{\mu}q)}-2\ {d_5}\ {(p_{\mu}q)}-2\
{f_5}\ {(p_{\mu}q)}\big) \nonumber
\end{eqnarray}

\newpage

\begin{eqnarray}
{\hbox{Im}\tilde{F_n}}&=&\frac{\alpha}{2\pi}{f_K}\ \Big(4\
{a_1}\
{m_{\mu}}+2\ {a_3}\ {m_{\mu}}+2\ {b_1}\
{m_{\mu}}+{b_{11}}\ {m_{\mu}}-2\ {b_4}\
{m_{\mu}}-2\ {c_4}\ {m_{\mu}}-  \nonumber \\
&& {J_{12}}\ {m_{\mu}}-2\ {J_2}\
{m_{\mu}}-{b_2}\ {{{m_K}}^2}\
{m_{\mu}}+{g_3}\ {{{m_K}}^2}\
{m_{\mu}}-2\ {a_2}\ {{{m_{\mu}}}^3}-  \nonumber \\
&& {c_2}\ {{{m_{\mu}}}^3}+{c_3}\
{{{m_{\mu}}}^3}+{f_3}\ {{{m_{\mu}}}^3}-2\
{a_2}\ {m_{\mu}}\ {(p_Kp_{\mu})}-2\
{b_2}\ {m_{\mu}}\ {(p_Kp_{\mu})}+  \nonumber \\
&& 2\ {b_3}\ {m_{\mu}}\
{(p_Kp_{\mu})}-2\ {c_2}\ {m_{\mu}}\
{(p_Kp_{\mu})}+2\ {d_3}\ {m_{\mu}}\
{(p_Kp_{\mu})}+2\ {J_1}\ {m_{\mu}}\
{(p_Kp_{\mu})}+ \nonumber  \\
&& 2\ {b_3}\ {m_{\mu}}\
{(p_Kq)}-\frac{{a_{12}}\
{{{m_{\mu}}}^3}}{{{{(p_{\mu}q)}}^2}}-\frac{{J_{11
}}\
{{{m_{\mu}}}^3}}{{{{(p_{\mu}q)}}^2}}-\frac{{a_{12
}}\ {m_{\mu}}}{{(p_{\mu}q)}}-\frac{2\ {a_4}\
{m_{\mu}}}{{(p_{\mu}q)}}+  \nonumber \\
&& \frac{{J_{11}}\
{m_{\mu}}}{{(p_{\mu}q)}}-\frac{{a_{11}}\
{{{m_K}}^2}\ {m_{\mu}}}{2\ {(p_{\mu}q)}}+\frac{3\
{a_1}\ {{{m_{\mu}}}^3}}{{(p_{\mu}q)}}+\frac{3\
{b_1}\
{{{m_{\mu}}}^3}}{{(p_{\mu}q)}}+\frac{{b_{11}}\
{{{m_{\mu}}}^3}}{2\ {(p_{\mu}q)}}-  \nonumber \\
&& \frac{2\ {J_2}\
{{{m_{\mu}}}^3}}{{(p_{\mu}q)}}-\frac{{b_{11}}\
{m_{\mu}}\
{(p_Kp_{\mu})}}{{(p_{\mu}q)}}+\frac{{J_{12}}\
{m_{\mu}}\
{(p_Kp_{\mu})}}{{(p_{\mu}q)}}-\frac{{b_{11}}\
{m_{\mu}}\ {(p_Kq)}}{{(p_{\mu}q)}}+ \nonumber  \\
&& \frac{{J_{12}}\ {m_{\mu}}\
{(p_Kq)}}{{(p_{\mu}q)}}-2\ {a_2}\
{m_{\mu}}\ {(p_{\mu}q)}+2\ {c_3}\
{m_{\mu}}\ {(p_{\mu}q)}+2\ {f_3}\
{m_{\mu}}\ {(p_{\mu}q)}\Big)+  \nonumber \\
&&\frac{\alpha}{2\pi} {F_v}\ \Big(2\ {a_4}\
{m_{\mu}}-4\
{a_5}\ {m_{\mu}}+2\ {b_{13}}\
{m_{\mu}}-4\ {b_5}\ {m_{\mu}}-  \nonumber \\
&& 2\ {a_1}\ {{{m_{\mu}}}^3}+{c_4}\
{{{m_{\mu}}}^3}-{c_5}\
{{{m_{\mu}}}^3}-{d_4}\
{{{m_{\mu}}}^3}-{d_5}\ {{{m_{\mu}}}^3}-3\
{e_5}\ {{{m_{\mu}}}^3}-  \nonumber \\
&& 3\ {f_5}\ {{{m_{\mu}}}^3}+2\ {a_1}\
{m_{\mu}}\ {(p_Kp_{\mu})}-2\ {c_4}\
{m_{\mu}}\ {(p_Kp_{\mu})}+2\ {d_4}\
{m_{\mu}}\ {(p_Kp_{\mu})}+  \nonumber \\
&& 2\ {d_5}\ {m_{\mu}}\
{(p_Kp_{\mu})}+2\ {e_5}\ {m_{\mu}}\
{(p_Kp_{\mu})}+4\ {f_5}\ {m_{\mu}}\
{(p_Kp_{\mu})}-2\ {c_4}\ {m_{\mu}}\
{(p_Kq)}+  \nonumber \\
&& 2\ {d_5}\ {m_{\mu}}\ {(p_Kq)}+2\
{f_5}\ {m_{\mu}}\ {(p_Kq)}+\frac{3\
{a_{13}}\
{m_{\mu}}}{{(p_{\mu}q)}}+\frac{{b_{13}}\
{{{m_{\mu}}}^3}}{{(p_{\mu}q)}}- \nonumber  \\
&& \frac{{b_{13}}\ {m_{\mu}}\
{(p_Kp_{\mu})}}{{(p_{\mu}q)}}-\frac{{b_{13}}\
{m_{\mu}}\ {(p_Kq)}}{{(p_{\mu}q)}}-2\
{a_1}\ {m_{\mu}}\ {(p_{\mu}q)}+2\
{c_4}\ {m_{\mu}}\ {(p_{\mu}q)}-  \nonumber \\
&& 2\ {d_4}\ {m_{\mu}}\
{(p_{\mu}q)}-2\ {d_5}\ {m_{\mu}}\
{(p_{\mu}q)}-2\ {e_5}\ {m_{\mu}}\
{(p_{\mu}q)}-4\ {f_5}\ {m_{\mu}}\
{(p_{\mu}q)}\Big)+  \nonumber \\
&& \frac{\alpha}{2\pi}{F_a}\ \Big(-6\ {a_4}\
{m_{\mu}}+8\
{a_5}\ {m_{\mu}}-{b_{13}}\ {m_{\mu}}+8\
{b_5}\ {m_{\mu}}-4\ {b_4}\
{{{m_{\mu}}}^3}- \nonumber  \\
&& 2\ {c_4}\ {{{m_{\mu}}}^3}+{c_5}\
{{{m_{\mu}}}^3}-2\ {d_4}\
{{{m_{\mu}}}^3}+{d_5}\ {{{m_{\mu}}}^3}+3\
{e_5}\ {{{m_{\mu}}}^3}+3\ {f_5}\
{{{m_{\mu}}}^3}+  \nonumber \\
&& 2\ {a_1}\ {m_{\mu}}\
{(p_Kp_{\mu})}-2\ {c_4}\ {m_{\mu}}\
{(p_Kp_{\mu})}+2\ {d_4}\ {m_{\mu}}\
{(p_Kp_{\mu})}+2\ {d_5}\ {m_{\mu}}\
{(p_Kp_{\mu})}+ \nonumber  \\
&& 2\ {e_5}\ {m_{\mu}}\
{(p_Kp_{\mu})}+4\ {f_5}\ {m_{\mu}}\
{(p_Kp_{\mu})}-2\ {c_4}\ {m_{\mu}}\
{(p_Kq)}+2\ {d_5}\ {m_{\mu}}\ {(p_Kq)}+
\nonumber \\
&& 2\ {f_5}\ {m_{\mu}}\
{(p_Kq)}-\frac{3\ {a_{13}}\
{m_{\mu}}}{{(p_{\mu}q)}}-\frac{{b_{13}}\
{{{m_{\mu}}}^3}}{2\ {(p_{\mu}q)}}-\frac{{b_{13}}\
{m_{\mu}}\ {(p_Kp_{\mu})}}{{(p_{\mu}q)}}-  \nonumber \\
&& \frac{{b_{13}}\ {m_{\mu}}\
{(p_Kq)}}{{(p_{\mu}q)}}-6\ {b_4}\
{m_{\mu}}\ {(p_{\mu}q)}-4\ {c_4}\
{m_{\mu}}\ {(p_{\mu}q)}-2\ {d_4}\
{m_{\mu}}\ {(p_{\mu}q)}+  \nonumber \\
&& 2\ {d_5}\ {m_{\mu}}\
{(p_{\mu}q)}+2\ {e_5}\ {m_{\mu}}\
{(p_{\mu}q)}+4\ {f_5}\ {m_{\mu}}\
{(p_{\mu}q)}\Big)\nonumber 
\end{eqnarray}

\newpage

\begin{eqnarray}
{\hbox{Im}\tilde{F_v}}&=&\frac{\alpha}{2\pi}{f_K}\
\Big({a_2}\
{{{m_{\mu}}}^2}+{c_3}\
{{{m_{\mu}}}^2}+{e_3}\ {{{m_{\mu}}}^2}+ \nonumber  \\
&& \frac{{a_1}\
{{{m_{\mu}}}^2}}{{(p_{\mu}q)}}+\frac{{b_1}\
{{{m_{\mu}}}^2}}{{(p_{\mu}q)}}-\frac{2\ {b_4}\
{{{m_{\mu}}}^2}}{{(p_{\mu}q)}}-\frac{{c_4}\
{{{m_{\mu}}}^2}}{{(p_{\mu}q)}}-\frac{{d_4}\
{{{m_{\mu}}}^2}}{{(p_{\mu}q)}}\Big)+ \nonumber  \\
&& \frac{\alpha}{2\pi}{F_a}\ \big(6\ {a_4}-2\
{a_5}-8\
{b_5}+{c_4}\ {{{m_{\mu}}}^2}-{d_4}\
{{{m_{\mu}}}^2}-{d_5}\
{{{m_{\mu}}}^2}-{e_5}\ {{{m_{\mu}}}^2}-  \nonumber \\
&& 2\ {f_5}\ {{{m_{\mu}}}^2}-2\ {a_1}\
{(p_Kp_{\mu})}+4\ {b_4}\ {(p_Kp_{\mu})}+4\
{c_4}\ {(p_Kp_{\mu})}-2\ {d_5}\
{(p_Kp_{\mu})}- \nonumber  \\
&& 2\ {e_5}\ {(p_Kp_{\mu})}-4\ {f_5}\
{(p_Kp_{\mu})}-2\ {a_1}\ {(p_Kq)}+2\
{b_4}\ {(p_Kq)}+4\ {c_4}\ {(p_Kq)}-  \nonumber \\
&& 2\ {d_5}\ {(p_Kq)}-2\ {f_5}\
{(p_Kq)}+2\ {c_4}\ {(p_{\mu}q)}-2\
{d_5}\ {(p_{\mu}q)}-2\ {f_5}\
{(p_{\mu}q)}\big)+  \nonumber \\
&& \frac{\alpha}{2\pi}{F_v}\ \big(-8\ {a_4}+4\
{a_5}+4\
{b_5}+2\ {a_1}\ {{{m_{\mu}}}^2}-4\
{b_4}\ {{{m_{\mu}}}^2}-3\ {c_4}\
{{{m_{\mu}}}^2}+{c_5}\ {{{m_{\mu}}}^2}-  \nonumber \\
&& {d_4}\ {{{m_{\mu}}}^2}+{d_5}\
{{{m_{\mu}}}^2}+3\ {e_5}\ {{{m_{\mu}}}^2}+3\
{f_5}\ {{{m_{\mu}}}^2}-2\ {a_1}\
{(p_Kp_{\mu})}+4\ {b_4}\ {(p_Kp_{\mu})}+ \nonumber  \\
&& 4\ {c_4}\ {(p_Kp_{\mu})}-2\ {d_5}\
{(p_Kp_{\mu})}-2\ {e_5}\ {(p_Kp_{\mu})}-4\
{f_5}\ {(p_Kp_{\mu})}-2\ {a_1}\
{(p_Kq)}+ \nonumber  \\
&& 2\ {b_4}\ {(p_Kq)}+4\ {c_4}\
{(p_Kq)}-2\ {d_5}\ {(p_Kq)}-2\ {f_5}\
{(p_Kq)}+4\ {a_1}\ {(p_{\mu}q)}-  \nonumber \\
&& 6\ {b_4}\ {(p_{\mu}q)}-6\ {c_4}\
{(p_{\mu}q)}+2\ {d_5}\ {(p_{\mu}q)}+2\
{e_5}\ {(p_{\mu}q)}+4\ {f_5}\
{(p_{\mu}q)}\big) \nonumber 
\end{eqnarray}

\newpage
\section*{Figure captions}

\vspace*{1.5cm}
\noindent
\begin{description}
\item[Fig. 1.] Feynman diagrams for the  $K^{\pm}\to \mu^{\pm}\nu\gamma$ decay
in the tree level of SM.
\item[Fig. 2.] Feynman diagrams contributing to
the muon transverse polarization in the tree level approximation of SM.
\item[Fig. 3.] Bounds for the three doublets Weinberg model parameters 
coming from: а) the analog of the KM matrix for the charged-Higgs boson mixings.
The allowed parameter region is below the bounding curve;
б)  the data  on charged-Higgs search at LEP~II (allowed region above the dashed
line) and data on the $ D^0-{\bar D}^0 $-mixing (allowed region is above
the solid line);
в) the data on the neutron EDM (allowed region is below the bounding curve).
\item[Fig. 4.]
The 3D Dalitz plot for the muon transverse polarization
as a function of  $x=2E_\gamma /m_K$ and $y=2E_\mu /m_K $ for the one-loop
approximation of SM.
\item[Fig. 5.] Level lines for the Dalitz plot of the muon transverse
polarization $P_T=f(x,y)$ in the SM case.
\item[Fig. 6.]
The 3D Dalitz plot for the muon transverse polarization
as a function of  $x=2E_\gamma /m_K$ and $y=2E_\mu /m_K $ within  the
three doublets Weinberg model.
\item[Fig. 7.] 
Level lines for the Dalitz plot of the muon transverse
polarization $P_T=f(x,y)$ in the Weinberg model case.
\end{description}

\newpage
\setlength{\unitlength}{1mm}
\begin{figure}[ph]
\bf
\begin{picture}(150, 200)

\put(10,160){\epsfxsize=10cm \epsfbox{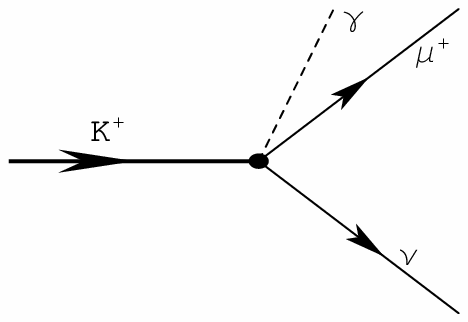}}
\put(28,160){Fig. 1a}

\put(80,160){\epsfxsize=10cm \epsfbox{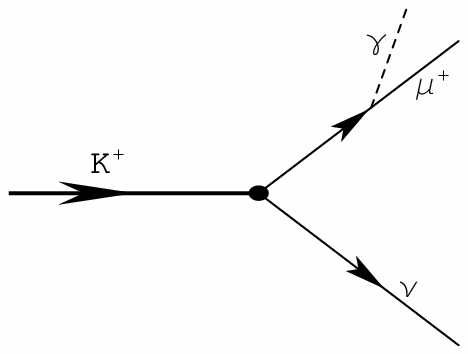}}
\put(98,160){Fig. 1b}

\put(45,60){\epsfxsize=10cm \epsfbox{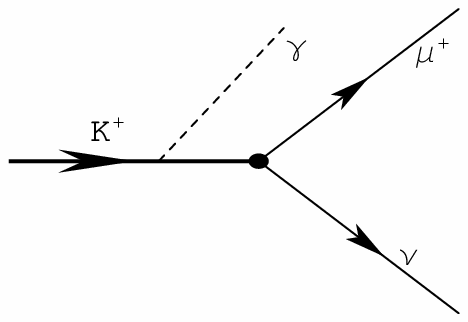}}
\put(64,60){Fig. 1c}

\end{picture}
\end{figure}

\newpage
\setlength{\unitlength}{1mm}
\begin{figure}[ph]
\bf
\begin{picture}(150, 200)

\put(5,170){\epsfxsize=8cm \epsfbox{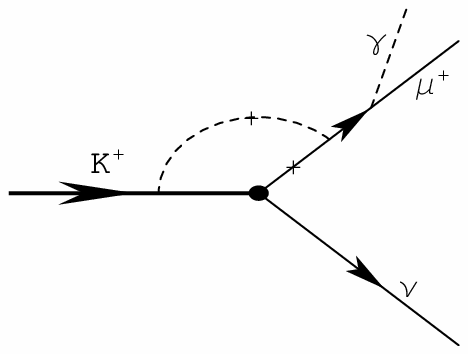}}
\put(23,175){Fig. 2a}

\put(95,170){\epsfxsize=8cm \epsfbox{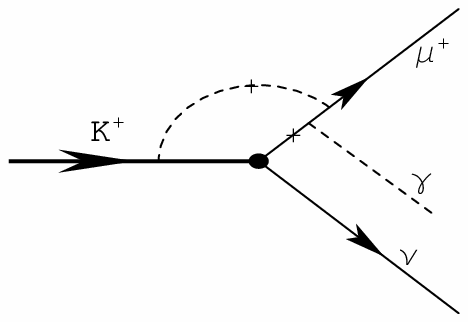}}
\put(123,175){Fig. 2b}

\put(5,110){\epsfxsize=8cm \epsfbox{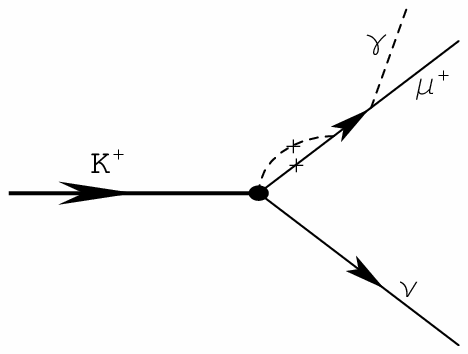}}
\put(23,115){Fig. 2c}

\put(95,110){\epsfxsize=9cm \epsfbox{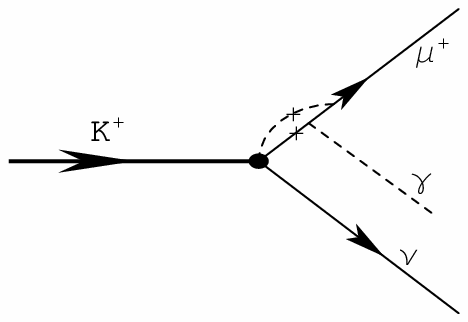}}
\put(123,115){Fig. 2d}

\put(5,50){\epsfxsize=8cm \epsfbox{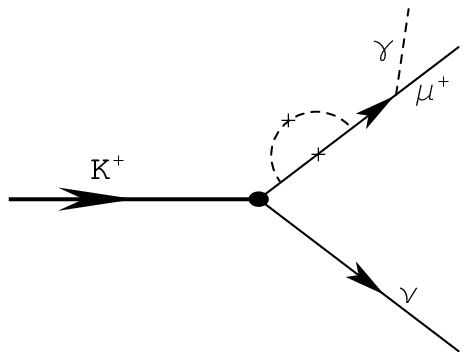}}
\put(23,55){Fig. 2e}

\put(95,50){\epsfxsize=9cm \epsfbox{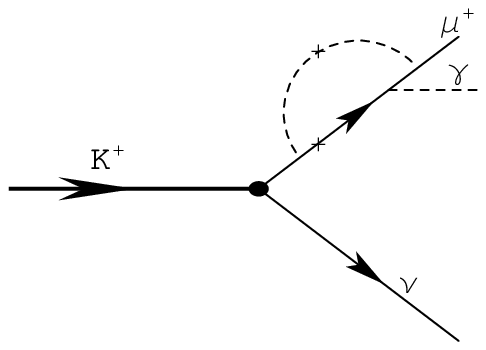}}
\put(123,55){Fig. 2f}

\put(50,0){\epsfxsize=9cm \epsfbox{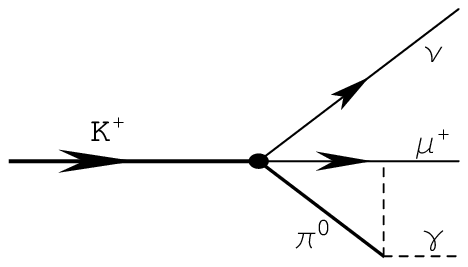}}
\put(73,5){Fig. 2g}

\end{picture}
\end{figure}

\newpage
\setlength{\unitlength}{1mm}
\begin{figure}[ph]
\bf
\begin{picture}(150, 200)

\put(-30,20){\epsfxsize=14cm \epsfbox{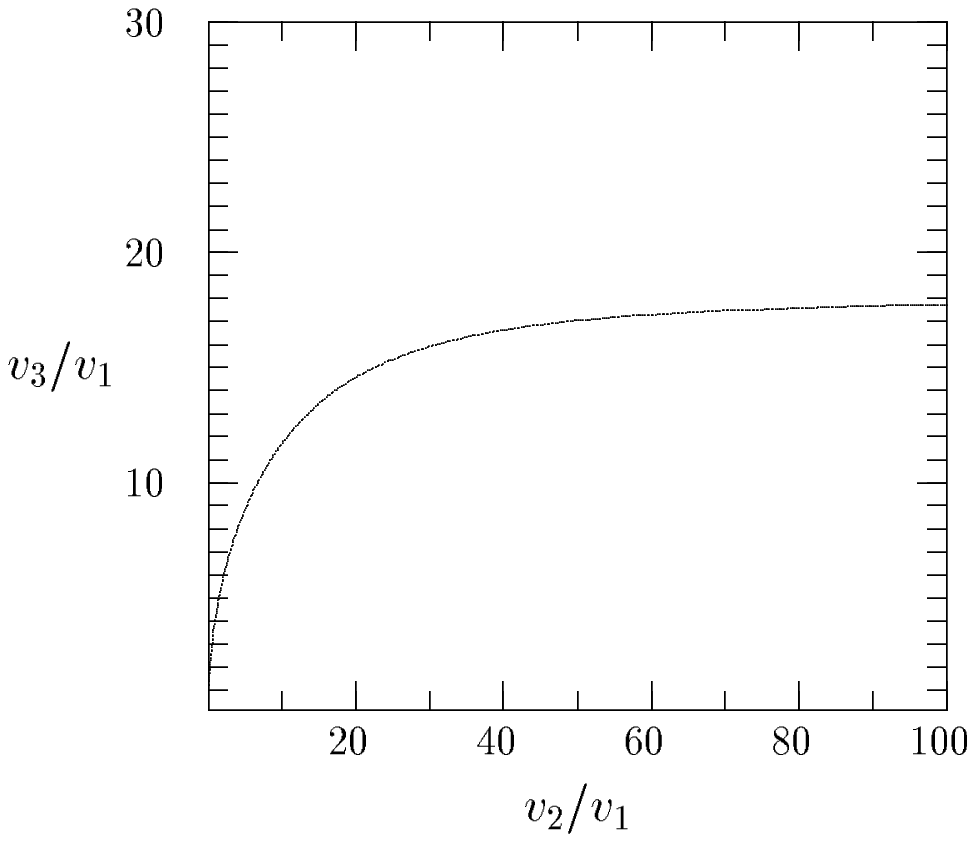}}
\put(38,100){Fig. 3a}

\put(50,20){\epsfxsize=14cm \epsfbox{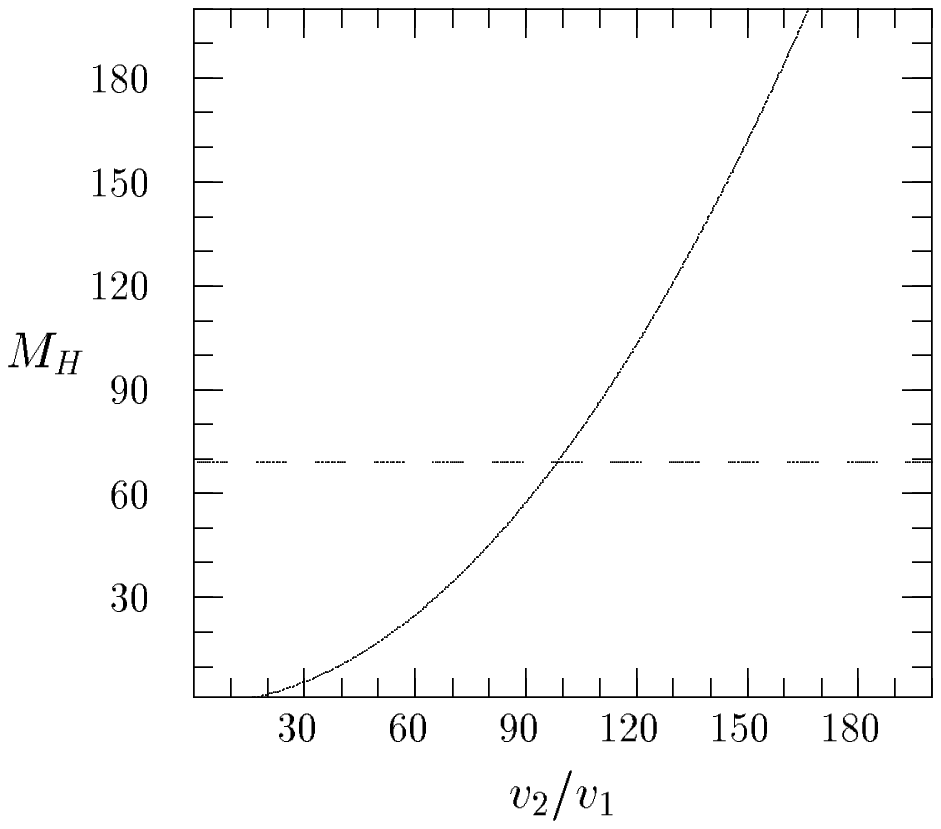}}
\put(108,100){Fig. 3b}

\put(10,-80){\epsfxsize=14cm \epsfbox{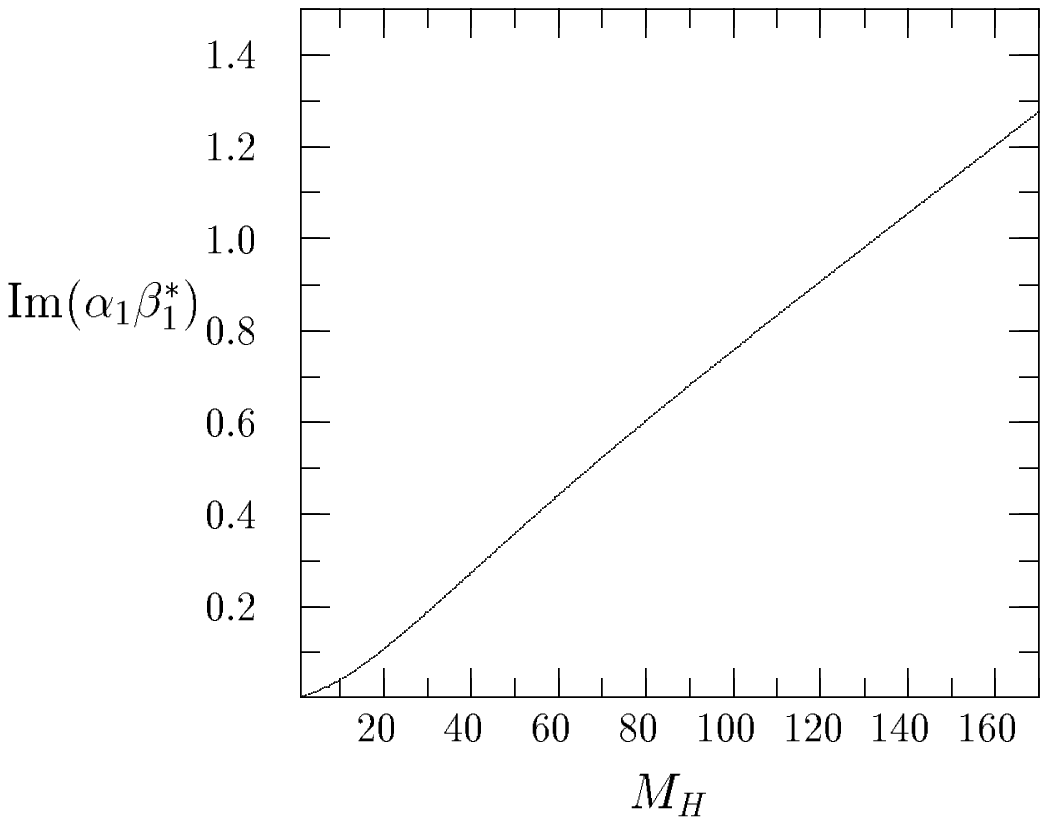}}
\put(74,0){Fig. 3c}

\end{picture}
\end{figure}

\newpage
\setlength{\unitlength}{1mm}
\begin{figure}[ph]
\bf
\begin{picture}(150, 200)
\put(40,130){\epsfxsize=9cm \epsfbox{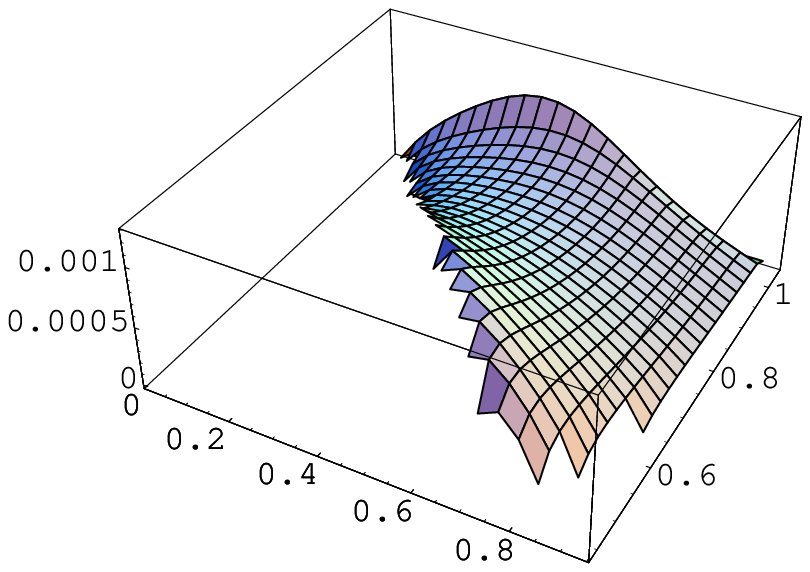}}
\put(80,120){Fig. 6}
\put(75,133){$\bf x$}
\put(125,145){$\bf y$}
\put(35,180){$\bf P_T$}

\put(40,25){\epsfxsize=10cm \epsfbox{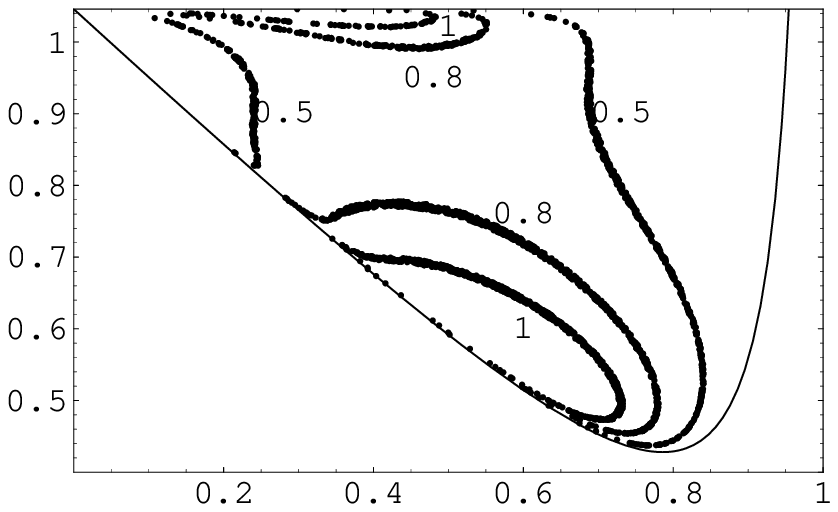}}
\put(90,90){$ P_T\cdot 10^{3}$}
\put(95,21){$\bf x$}
\put(35,60){$\bf y$}

\put(80,10){Fig. 5}

\end{picture}
\end{figure}

\newpage
\setlength{\unitlength}{1mm}
\begin{figure}[ph]
\bf
\begin{picture}(150, 200)
\put(40,130){\epsfxsize=9cm \epsfbox{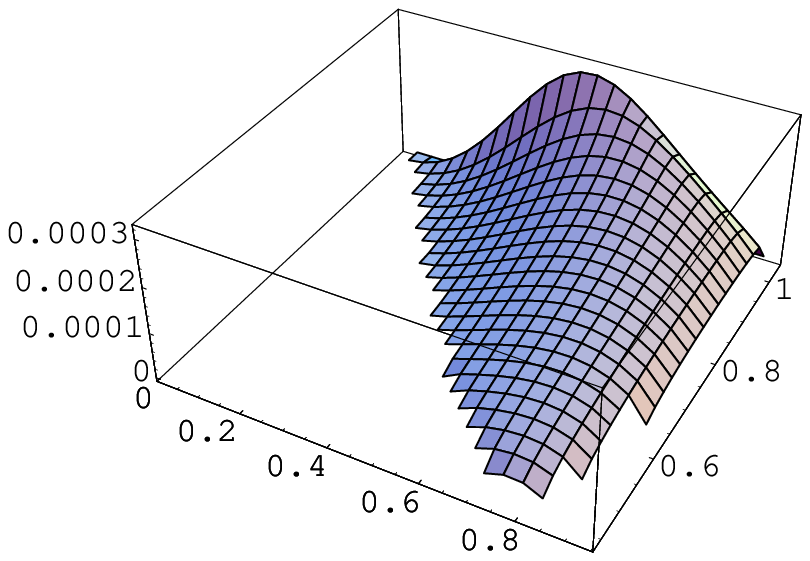}}
\put(80,120){Fig. 6}
\put(75,133){$\bf x$}
\put(125,145){$\bf y$}
\put(35,180){$\bf -P_T$}

\put(0,-100){\epsfxsize=18cm \epsfbox{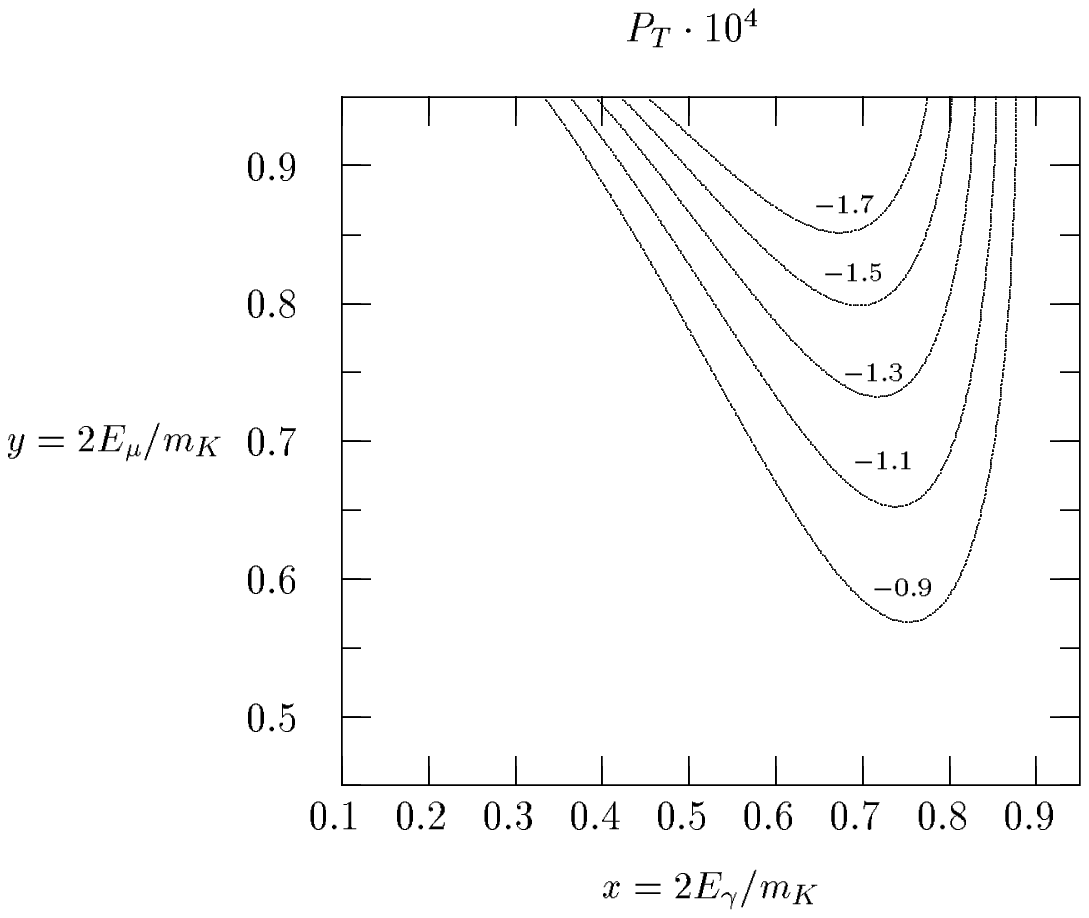}}

\put(80,10){Fig. 7}

\end{picture}
\end{figure}

\end{document}